\documentclass[fleqn,twoside,twocolumn,nofootinbib,showkeys]{revtex4} 
\usepackage[sec,nocpr]{ujp} 

\begin{document}
\title[Temperature Effects]%
{TEMPERATURE EFFECTS ON THE SURFACE PLASMON RESONANCE IN COPPER NANOPARTICLES}%
\author{O.A.~Yeshchenko}
\affiliation{Faculty of Physics, Taras Shevchenko National University of Kyiv}%
\address{4, Prosp. Academician Glushkov, Kyiv 03127, Ukraine}%
\email{yes@univ.kiev.ua}

\udk{539.2; 535.34} \pacs{73.20.Mf, 78.67.Bf,\\[-3pt] 65.80.-g}
\razd{\secviii}

\autorcol{O.A.\hspace*{0.7mm}Yeshchenko}

\setcounter{page}{249}%
\begin{abstract}
The temperature dependences of the energy and the width of a surface
plasmon resonance are studied for copper nanoparticles
17--59~nm in size in the silica host matrix in the temperature interval 293--460~K.
An increase of the temperature leads to the red shift and the broadening of the
surface plasmon resonance in Cu nanoparticles. The obtained
dependences are analyzed within the framework of a theoretical model
considering the thermal expansion of a nanoparticle, the
electron-phonon scattering in a nanoparticle, and the temperature
dependence of the dielectric permittivity of the host matrix. The
thermal expansion is shown to be the main mechanism responsible for
the temperature-induced red shift of the surface plasmon resonance
in copper nanoparticles. The thermal volume expansion coefficient
for Cu nanoparticles is found to be size-independent in the studied
size range. Meanwhile, the increase of the electron-phonon scattering
rate with the temperature is shown to be the dominant
mechanism of the surface plasmon resonance broadening in copper
nanoparticles.
\end{abstract}

\maketitle

\section{Introduction}

Noble metal nanoparticles exhibit unique optical properties such as
the resonant absorption and the scattering of light, which are not found in bulk
counterparts \cite{1,2}. Collective coherent oscillations of the
free electrons in the conduction band, also known as a Surface Plasmon
(SP) resonance, are responsible for the strong absorption
and the scattering of light by particles \cite{1}. The energy and
the width of an SP resonance depend on the size, morphology, spatial
orientation, and optical constants of particles and the embedding
medium \cite{1,2}. The excitation of an SP resonance in a metal
nanoparticle leads to an enhanced local electric field close to the
surface of the particle \cite{1,2,3,4}. Therefore, noble metal
nanoparticles have attracted a lot of attention recently due to a
wide range of potential applications in the surface enhanced Raman
scattering (SERS) \cite{5,6,7}, surface enhanced fluorescence
\cite{8,9,10}, biochemical imaging \cite{11,12,13}, cancer
treatment \cite{11,14,15,16}, and subwavelength optical waveguides
\cite{17,18,19,20,21}, to name just a few.

An influence of the temperature on the SP resonance in metal
nanoparticles is crucial for the pure and applied science of
nanoparticles \cite{1,22}. The temperature dependence of the SP
resonance is important because of the recent applications of noble
metal nanoparticles to the thermally assisted magnetic recording
\cite{23}, thermal cancer treatment \cite{16,24,25,26} catalysis and
nanostructure growth \cite{27}, and computer chips \cite{28}.
However, the SP temperature dependence was not studied in detail to
date, because a broad temperature interval requires a usage of
materials with high thermal stability. Colloids, thin films, and
glasses, which are major media in optical materials with metal
nanoparticles, possess no enough thermal stability (800--900 K at
maximum for glasses). The more advanced material which is highly
transparent in the optical range is silica, and the facile sol-gel
technique allows one to produce metal nanoparticles of the different
chemical nature and the size range within the matrix that is stable
in air up to 1600 K. In the present work, we use the silica sol-gel
glasses admitting the usage of a wider temperature interval without
the sample destruction. For these samples, the full reversibility in
the variation of optical properties occurred.

The underlying physics of the temperature dependence of optical
properties of metal nanoparticles is a precondition for the
development of successful and reliable applications and devices. The
temperature effects for the SP resonance absorption band in metal
nanoparticles were studied, e.g., by Kreibig \cite{1,29}, Doremus
\cite{30,31}, and the origin of temperature effects upon the SP
resonance was analyzed by Mulvaney \cite{32}. Recently, the
influence of the temperature on the SP resonance in Au-based plasmonic
nanostructures was reported for low temperatures 80--400 K in
Ref. \cite{33}, where the appreciable red shift and the SP
resonance broadening with increasing temperature were observed. The
temperature-induced broadening of the SP extinction band leaded to an appreciable
decrease of the light exctinction at the SP resonance frequency and,
respectively, to an increase of the extinction on the wings of the SP
extinction band. The interplay of temperature effects and a material
quality in nanoscale plasmonic waveguiding structures was
discussed in Ref. \cite{34}. So, most of the effects related to the
temperature dependence of the SP resonance were observed for low
temperatures \cite{1,33}. Therefore, there is a lack of data on
the influence of the temperature on an SP resonance in metal nanostructures for
temperatures higher than room one. The
temperature dependence of the SP resonance energy and bandwidth for
gold colloidal nanoparticles was studied by Link and El-Sayed \cite{35} for temperatures higher than
room one. Gold colloidal nanoparticles within the size range from 9 nm
to 99 nm were studied at elevated temperatures up to 365 K. No
significant influence of the temperature on the SP resonance energy
and bandwidth was found. In our recent work \cite{36}, we
studied the temperature dependence of the SP resonance in silver
nanoparticles at high temperatures in the silica matrix similar to
that in the present work. We observed the noticeable red shift and
the broadening of the SP resonance in Ag nanoparticles with increasing
temperature, which is in full accordance with effects reported for
plasmonic Au-nanostructures at low temperatures.

In this paper, we present experimental results on the temperature
dependence of the SP resonance energy and width in copper
nanoparticles 17--59 nm in size embedded in the
silica matrix in the temperature interval 293--460 K. We have
observed that an increase of the temperature of a sample leads to
the red shift and the broadening of the SP resonance, which is similar to our
results \cite{36} obtained for Ag nanoparticles in silica at high
temperatures and to results of Bouillard {\it et al.} \cite{33}
obtained for Au nanostructures at low temperatures. We analyze the
observed temperature dependence of the SP resonance within the framework
of a theoretical model considering such phenomena as the thermal volume
expansion of a nanoparticle, electron-phonon scattering in a
nanoparticle, and temperature dependence of the dielectric permittivity
of the host matrix. We show that the main mechanism of the red shift of
the SP resonance with increasing temperature is the thermal volume
expansion of a nanoparticle, while the electron-phonon scattering
in a nanoparticle is the dominant mechanism of SP resonance
broadening. The paper is organized as follows. After the introduction,
we give a short description of the procedure of synthesis of
copper nanoparticles and a structural and optical characterization of
nanoparticles. In Section 3, we describe the experimental
observations of the red shift and the broadening of the SP resonance, as
the temperature increases. In Section 4, we outline a theoretical
model of the temperature dependence of the SP resonance energy and
width. In Section 5, we give the comparative analysis of
experimental data and results of theoretical calculations and
discuss the significance of different mechanisms leading to the
temperature-induced red shift and the broadening of the SP resonance in
copper nanoparticles. In Conclusions, we summarize the main obtained
results.

\section{Synthesis of Copper Nanoparticles and~Their Optical and Structural Characterization}

Composite Cu/SiO$_{2}$ samples with copper nanoparticles
were produced, by using the modified sol-gel technique based on
hydrolysis of tetraethoxysilane (TEOS) with Cu-doping followed by a
chemical transformation of the precursors of dopants under annealing in
air or controlled gaseous medium. A precursor sol was prepared by
the mixing of TEOS, water, and ethyl alcohol with the acid catalyst such as
HNO$_{3}$ or HCl. The silica powder with a particle size of about 5--15 nm
(the specific surface area is 380 $\pm$ 30 m$^{2}$/g) was added to
the sol followed by ultrasonication in order to prevent a large
volume contraction during drying. The next gelation step resulted in
the formation of gels from sols. Porous materials (xerogels)
were obtained, after the gels were dried at room temperature.
The porosity of SiO$_{2}$ matrices was controlled by annealing the
samples in air at $600 ~^\circ$C during 1~h. Doping by copper was
performed by immersion of xerogels into a Cu(NO$_{3}$)$_{2}$ alcohol
solution during 24~h. Then the impregnated samples have been dried
in air at $40~^\circ$C during 24~h. The further processing of the
Cu-doped xerogels was done by two pathways. (1)~The initial
annealing in air with gradual increase of the temperature from
$20~^\circ$C to $1200~^\circ$C (annealing time at $1200~^\circ$C: 5
min), then the annealing in the atmosphere of molecular hydrogen at a
temperature of $800~^\circ$C during 1~h. (2)~An annealing in H$_{2}$
with gradual increase of the temperature from $20~^\circ$C to
$1200~^\circ$C (annealing time at 1200 $^{0}$C: 5 min). The
annealing resulted in the decomposition of Cu(NO$_{3}$)$_{2}$
followed by the nucleation and the aggregation of Cu clusters resulting in
the formation of Cu nanoparticles of various sizes. Glass samples
fabricated were polished up to a thickness of about 1~mm for optical
measurements. The sets of samples were prepared. The samples
obtained at the successive annealing in air and hydrogen are light-pink
colored (samples AH); ones obtained at annealing in hydrogen are red
colored (samples H).

The copper nanoparticles formed were characterized with transmission
electron microscopy (TEM) to determine their mean size and
morphology. TEM characterization was performed with the use a JEOL
\mbox{JEM-2000EX} electron microscope. Figure 1 shows the typical
TEM micrographs of some samples. It can be seen that TEM indicates a
large separation between nanoparticles. Therefore, the
electrodynamical coupling cannot affect their optical spectra, and
we use one-particle models for simulations below. The micrographs
prove the creation of spherical Cu nanoparticles of various
sizes in the matrix in dependence on the annealing conditions.
Nanoparticles with mean sizes of 17 nm and 35 nm are formed in
the samples annealed successively in air and then in molecular
hydrogen (samples AH). Cu nanoparticles with mean sizes of 48 nm
and 59 nm are formed in the samples annealed in H$_{2}$ (samples
H). TEM shows that Cu nanoparticles in the studied samples are
characterized by the Gaussian size distribution with clear-featured
maximum. The dispersion of the size distribution is quite low in all the
samples being in range of 11--17 $\%$ depending on the sample. Since
the distribution dispersion is small, it is reasonable to conclude that
the effect of the particle size distribution would not affect the shape and
the width of the SP absorption band of copper nanoparticles.

A tungsten-halogen incandescent lamp was used as a light source in
the measurements of absorption spectra. A single grating
spectrometer MDR--3 was used for the registration of spectra.
The samples were placed into an open furnace during the spectral
measurements. Each spectrum was measured at its own stabilized
temperature.

\begin{figure}
\includegraphics[width=5.0cm]{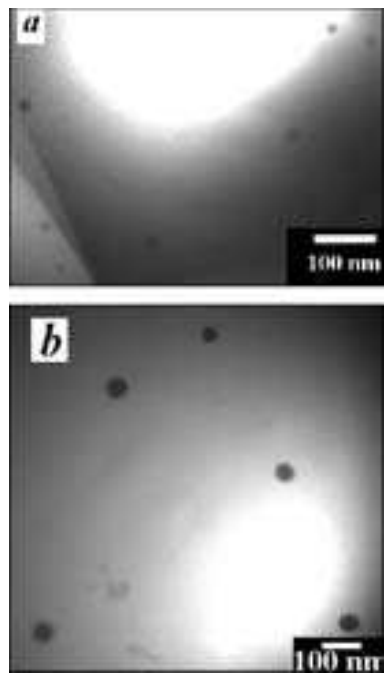} \caption{Typical TEM
micrographs of Cu nanoparticles in the silica matrix. {\it a} -- AH2
sample annealed successively in air and hydrogen ($ \langle d\rangle
= 17$~nm), {\it b}~-- H1 sample annealed in hydrogen ($  \langle
d\rangle = 48$~nm)}
\end{figure}

\begin{figure}
\vspace*{1mm}
\includegraphics[width=\column]{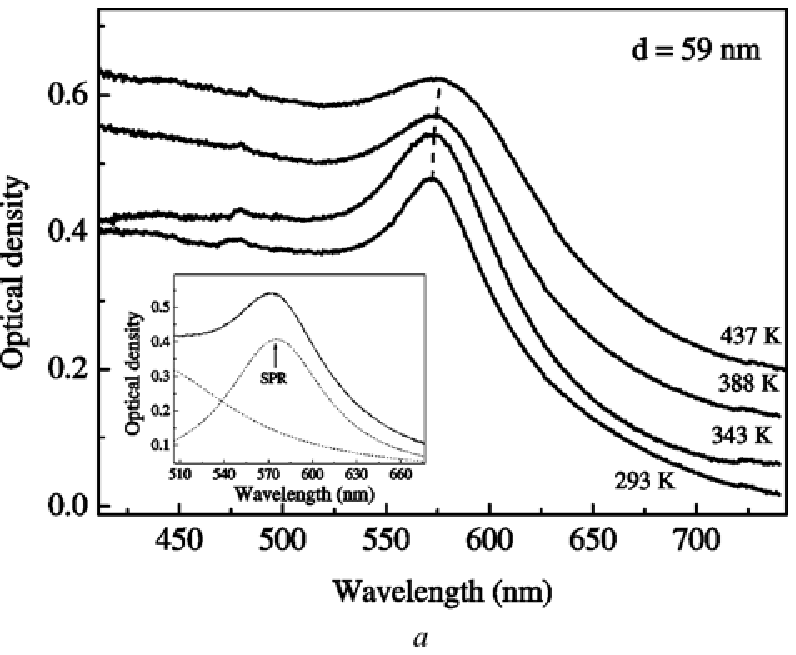}\\ [2mm]
\includegraphics[width=\column]{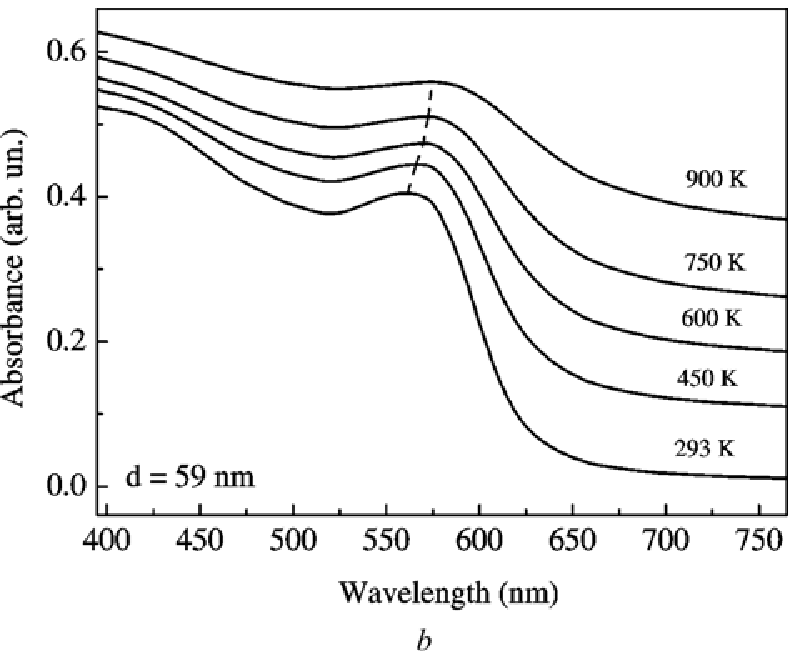}
\vskip-3mm\caption{Measured evolution of the absorption spectrum of
Cu nanoparticles with a mean size of 59 nm in silica with gradual
increase of the temperature from 293 to 437 K ({\it a}). Inset
illustrates the decomposition of the total spectrum into basic
spectral Lorentzian contours, where SPR is the SP resonance band in
a Cu nanoparticle. Calculated evolution of the absorption spectrum
of a 59-nm Cu nanoparticle in silica with increasing
temperature~({\it b})}
\end{figure}

\section{Temperature Dependence of~Surface~Plasmon Resonance in~Cu~Nanoparticles:~Experiment}

We measured the absorption spectra of copper nanoparticles in the silica
host matrix. Samples containing Cu nanoparticles with mean sizes of
17, 35, 48, and 59 nm were studied. Absorption
spectra were measured in the temperature interval 293--460 K. The aim
was to study the effects of the influence of the temperature on the
spectral characteristics (energy and bandwidth) of the surface plasmon
resonance in copper nanoparticles. Evolution of the experimental
absorption spectrum of the composite sample containing Cu
nanoparticles with a mean size of 59 nm is shown in Fig.~2, {\it a}.

We performed the decomposition of the total absorption spectra into
basic Lorentzian spectral contours. The studied absorption
spectra contain two bands. The high-energy band is caused by the
interband transitions in copper, and the low-energy one is caused by
the excitation of surface plasmons in Cu nanoparticles. For the studied
samples, the SP band is located in the spectral range 570 nm
(2.18 eV)--580 nm (2.14 eV) depending on the nanoparticle size and
the temperature. Such spectral position of the SP band is typical of copper
nanoparticles embedded in silica \cite{1}. Thus, the spectral
decomposition allowed us to extract the surface plasmon band from
the total spectrum and determine the spectral characteristics of
this band such as the spectral position (energy of SP resonance) and
the bandwidth (SP resonance width). The errors of the multipeak analysis
were small, specifically the error of determination of the SP energy
was about 0.5--0.7$\%$ and one of determination of the SP bandwidth
was within 5$\%$. Therefore, the SP band was extracted from the
total absorption spectrum quite accurately. This allowed us to
obtain width reliable experimental temperature dependences of the SP
resonance energy and bandwidth.

Figure 3 presents the obtained temperature dependences of the SP resonance
energy and bandwidth for copper nanoparticles of various sizes.
Figures 2 and 3 show that the monotonous increase of the temperature from
293 to 460 K leads to the monotonous red shift (shift to lower
frequencies) of the surface plasmon band and to its broadening. It is seen
that the manifestations of the temperature effects on the surface
plasmon resonance in copper nanoparticles are quite prominent. One
can see that the obtained dependences are not qualitatively different
for copper nanoparticles with different sizes in the studied
size range. Note that the observed temperature dependences for Cu
nanoparticles in silica are quite similar to ones observed for Ag
nanoparticles in the same host matrix \cite{36}, i.e. the red shift
and the broadening of the SP resonance with increasing temperature. We
note that we measured the absorption spectra of Cu/SiO$_{2}$
composite samples both at their heating and cooling. We observed the
full reversibility of the temperature behavior of the spectra. This
indicates the high thermal stability and the high optical quality of
sol-gel prepared Cu/SiO$_{2}$ nanocomposites. This is important for
applications of such nanocomposites in optical devices working
under extreme thermal conditions.

\section{Temperature Dependence of~Surface~Plasmon Resonance in~Metal~Nanoparticles:~Theory}

In this section, we give the theoretical analysis of various
mechanisms that can cause the observed temperature effects, namely
the red shift and the broadening of the SP resonance in metal (copper, in
particular) nanoparticles occurring with increasing
temperature. Those effects are: (1) electron-phonon scattering in
a nanoparticle, (2) thermal expansion of a nanoparticle and (3)
temperature dependence of the dielectric permittivity of the silica
host matrix.

It is well known (see, e.g., \cite{1,2}) that the ab\-sorp\-tion
coefficient of a composite with non-interacting
spherical metal nanoparticles much smaller than the light wavelength
($d \ll \lambda$) is
\begin{equation}
K \left( \omega \right) = \frac{9f \omega \varepsilon _{m}^{3/2}}{c}
\frac{\varepsilon _{2}}{\left( \varepsilon _{1} + 2\varepsilon _{m} \right)^2 + \varepsilon _{2}^{2}},
\label{eq1}%
\end{equation}
where $\omega$ is the frequency, $c$  is the light velocity,
$\varepsilon (\omega) = \varepsilon _{1} (\omega) + i\varepsilon
_{2} (\omega)$  is the dielectric permittivity of a nanoparticle,
$f$ is the filling factor of the composite, $\varepsilon _{m}$ is
the dielectric permittivity of the host matrix. It is clear that the
temperature dependence of the permittivities of a nanoparticle and
the host matrix would affect the energy and the width of the SP
resonance in a nanoparticle and, respectively, affect the absorption
spectrum of the composite. The dielectric permittivity of a metal
can be \mbox{expressed as}\looseness=1
\begin{equation}
\varepsilon (\omega) = \varepsilon _{ib} (\omega) + \varepsilon _{D} (\omega),
\label{eq2}%
\end{equation}

\noindent where $\varepsilon _{ib} (\omega) = \varepsilon _{ib1}
(\omega) +
 i \varepsilon _{ib2} (\omega)$ is the contribution of the interband transitions
(bound electrons) to the dielectric permittivity of a metal, and
$\varepsilon _{\rm D} (\omega)$  is the contribution of the free
electrons that is given by the Drude theory as
\begin{equation}
\varepsilon _{\rm D} (\omega) = 1 - \frac{\omega _{p}^{2}}{\omega
^{2} + i\gamma\omega}.
\label{eq3}%
\end{equation}

\begin{figure}
\vskip1mm
\includegraphics[width=5cm]{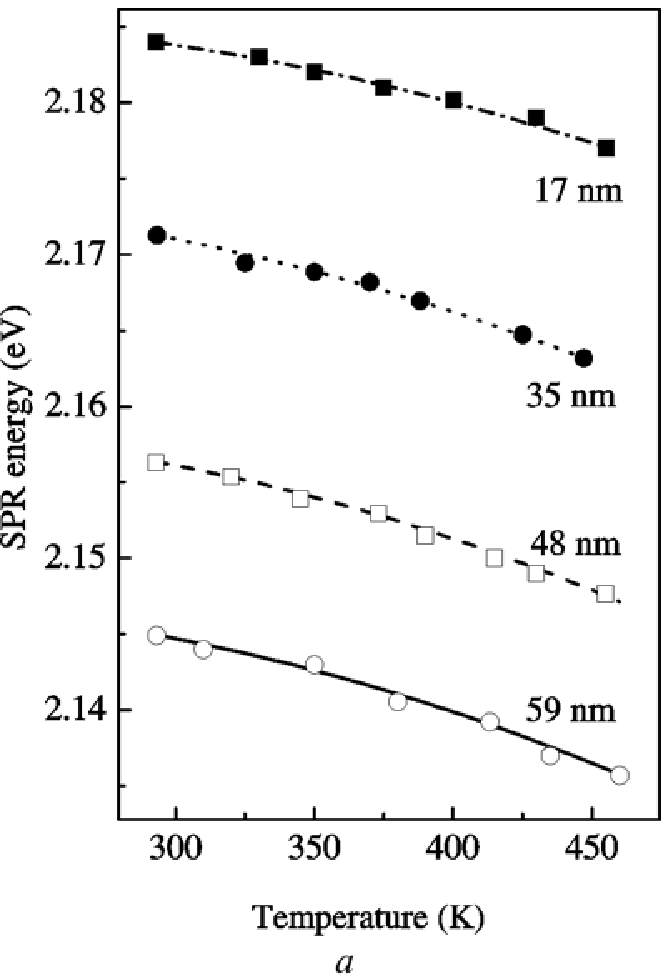}\\[2mm]
\includegraphics[width=5cm]{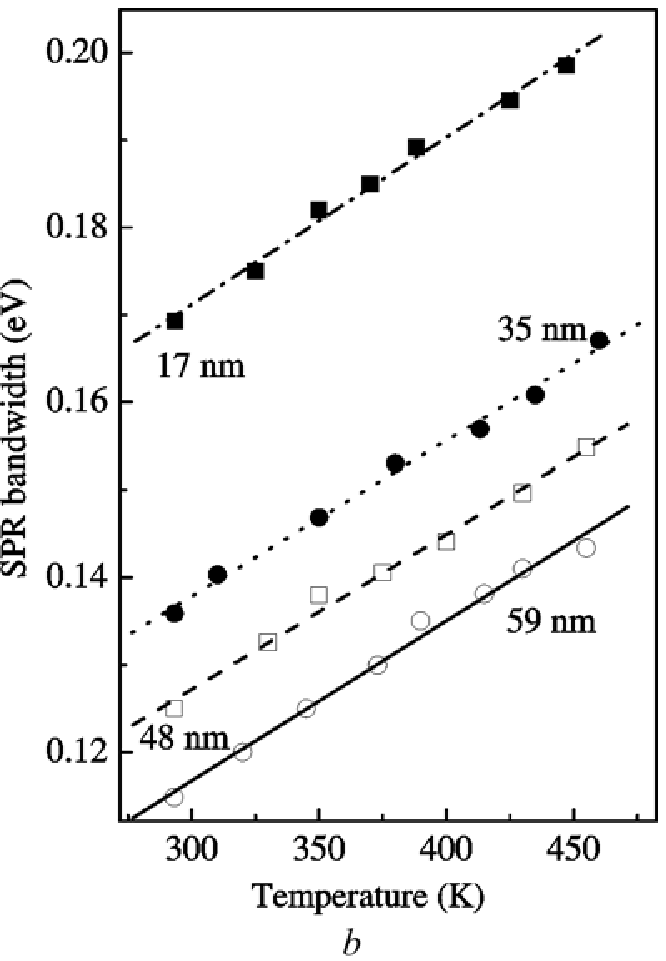}
\vskip-3mm\caption{Dependences of the SP resonance energy ({\it a})
and the bandwidth ({\it b}) for copper nanoparticles of various sizes in
silica. Points -- experiment, lines -- theory}
\end{figure}

\noindent Here,
\begin{equation}
 \omega _{p}  = \sqrt{\frac{4\pi n e^{2}}{m^{*}}}
\label{eq4}%
\end{equation}
is the bulk plasmon frequency, where $n$ is the concentration of free electrons,
$e$ is the electron charge, $m^{*}$  is the effective
mass of a free electron; $\gamma$ is the damping constant of plasma
oscillations. In the approximation of small damping ($\varepsilon_{2}
\ll | \varepsilon _{1} + 2\varepsilon _{m} |$), which is quite good
for noble metals, the condition of excitation of the SP resonance in
the nanoparticles much smaller than the light wavelength is as
follows \cite{1}:
\begin{equation}
\varepsilon _{1} (\omega) = -2\varepsilon _{m}.
\label{eq5}%
\end{equation}
The SP resonance energy is then obtained by substituting the complex
dielectric permittivity for a metal nanoparticle from Eqs. (\ref{eq2})
and (\ref{eq3}) into Eq. (\ref{eq5}):
\begin{equation}
\omega _{sp} = \sqrt{\frac{\omega _{p}^{2}}{1+\varepsilon_{ib1}+2\varepsilon_{m}}-\gamma^{2}}.
\label{eq6}%
\end{equation}
Here, $\varepsilon_{ib1}$  is the real part of the contribution of
interband transitions to the permittivity of a nanoparticle. The
plasmon damping constant can be expressed \cite{1,2}~as
\begin{equation}
\gamma = \gamma _{\infty} + A\frac{v_{\rm F}}{R},
\label{eq7}%
\end{equation}
where $R$ is the nanoparticle radius, $\gamma _{\infty}$ is the
size-independent damping constant caused by the scattering of free
electrons on electrons, phonons, and lattice defects, $A$ is a
theory-dependent parameter that includes details of the scattering
process (e.g., isotropic or diffuse scattering \cite{1,37,38}), and
$v_{\rm F}$  is the Fermi velocity in a bulk metal (1.57 $\times$
10$^{8}$ cm/s in bulk copper \cite{39}). Note, that there exists a
more correct theory of the size dependence of the surface plasmon
damping constant $\gamma (R)$ \cite{40}. This theory predicts the
oscillating character of $\gamma (R),$ which is a result of the
quasidiscrete spectrum of phonons in a nanoparticle. Really, the
electron scattering on long-wave phonons dominates in metal, and the
long-wave spectrum for metal nanoparticles is limited by the size of
a particle that has to lead to oscillations in the $\gamma (R)$
dependence. Note, however, that noticeable oscillations in the
dependence $\gamma (R)$ appear for the size of nanoparticles smaller
than about 15 nm. For larger particles, the oscillations are small
and decrease with increasing size. As a result, the dependence
$\gamma (R)$ for larger particles is nearly monotonically decreasing
and, respectively, is well described by expression (\ref{eq7}). The
nanoparticles studied in our work are quite large (17--59 nm), so
our choice of the simplified expression (\ref{eq7}) to describe the
size dependence of the surface plasmon damping constant is quite
correct. Moreover, since the nanoparticles in the samples under
study are not monosize, their size distribution has to lead to the
blurring of slight oscillations of the damping constant for large
\mbox{nanoparticles.}\looseness=1

Let us analyze the influence of the temperature on the energy and the width
of the SP resonance in a metal nanoparticle embedded in the host matrix. The first
mechanism of the dependence of the SP resonance on the temperature is
the electron-phonon scattering in a metal nanoparticle. The
size-independent damping constant $\gamma _{\infty}$ depends on
the temperature due to the temperature dependence of the electron-phonon
scattering rate. Really, at an increase of the temperature, the phonon
population in a metal increases, which leads to an increase of
the probability of the electron-phonon scattering, which results in the
increased scattering rate for electrons. The $\gamma _{\infty}
(T)$  dependence caused by the electron-phonon scattering is given by
\cite{41}
\begin{equation}
\gamma _{\infty} (T) = K^{'}T^{5} \int\limits_{0}^{\theta /
T}\frac{z^{4}dz}{e^{z}-1},
\label{eq8}%
\end{equation}
where $\theta$ = 343 K is the Debye temperature for copper \cite{39},
and $K^{'}$ is a constant \cite{41}. Knowing the bulk damping constant
$\gamma _{\infty}$ at a certain temperature (e.g., at room one,
$T_{0}$ = 293 K), $K^{'}$ can be calculated as
\begin{equation}
K^{'} = \gamma _{\infty} (T_{0}) / \left(\!T_{0}^{5}
\int\limits_{0}^{\theta / T_{0}}\frac{z^{4}dz}{e^{z}-1}\!\right)\!.
\label{eq9}%
\end{equation}
For bulk copper at $T_{0}$ = 293 K, we have $\gamma _{\infty}$ = 0.09 eV
\cite{42}. Thus, an increase of the electron-phonon scattering rate
with the temperature would lead to an increase of the damping
constant $\gamma _{\infty}$. Apparently, this would lead to the
SP resonance broadening and to its red shift in
accordance with Eq. (\ref{eq6}).

The second mechanism of the dependence $\varepsilon(T)$ is the thermal
expansion of a nanoparticle. Indeed,
the nanoparticle volume increases with the temperature,
\begin{equation}
V(T) = V_{0} \left( 1 + \beta \Delta T \right),
\label{eq10}%
\end{equation}
where $\Delta T = T - T_{0}$ is the change in the temperature from
room one ($T_{0}$ = 293 K), $\beta$ is the volume thermal expansion
coefficient (5.1$\times$10$^{-5}$ K$^{-1}$ for bulk copper
\cite{39}), $V_{0}$ is the volume of a nanoparticle at $T_{0}$ = 293
K. The density of free electrons in a metal particle is given by
$n = N/V$, where $N$ is the number of electrons, and $V$ is the
particle volume. We denote the free electron density at
room temperature by $n_{0}.$ Since the total number of free
electrons in a nanoparticle is temperature-independent \cite{43},
$N = n_{0}V_{0} = n(T)V(T)$, we combine Eqs. (\ref{eq4}) and
(\ref{eq10}) and obtain the expression for the frequency of a bulk
plasmon:
\begin{equation}
\omega _{p} = \sqrt{\frac{4\pi n_{0} e^{2}}{m^{*}\left( 1+ \beta \Delta T \right)}} .
\label{eq11}%
\end{equation}
Substituting Eq. (\ref{eq10}) in Eq. (\ref{eq6}), we obtain the
expression for the frequency of the SP resonance in a metal
nanoparticle
\begin{equation}
\omega_{sp}=\sqrt{\frac{\omega_{p0}^{2}}{\left(1+\varepsilon_{ib1}+2\varepsilon_{m}\right)
\left(1+\beta \Delta T \right)}-\gamma^{2}},
\label{eq12}%
\end{equation}
where $\omega_{p0}=\sqrt{4\pi n_{0}e^2/m^{*}}$ is the bulk plasmon
frequency at room temperature. Thus, the thermal expansion of
a nanoparticle would lead to a decrease of the concentration of
free electrons in a nanoparticle and, respectively, to a decrease
of the SP resonance energy, i.e. to its red shift with increasing
temperature. It is well known \cite{1} that the damping constant of
plasma oscillations depends on the size of a nanoparticle as
$\gamma(R)    \ \propto 1/R$ (see Eq. (\ref{eq7})). That is due to
the scattering of free electrons on the surface of a nanoparticle. Under
the thermal expansion, the radius of a nanoparticle increases as
\begin{equation}
R(T)=R_{0}\left(1+\beta\Delta T\right)^{1/3} \!,
\label{eq13}%
\end{equation}
where $R_{0}$ is the nanoparticle radius at room temperature.
Therefore, the thermal expansion of nanoparticles would affect the
SP resonance frequency not only through the frequency of a bulk
plasmon, but through the size-dependent part of the plasmon damping
constant as well. The volume expansion coefficient depends on the
temperature according \mbox{to \cite{44} as}\looseness=1
\begin{equation}
\beta(T)=\frac{192\rho k_{\rm B}}{r_{0}\phi \left(16\rho -7Tk_{\rm
B}\right)^2},
\label{eq14}%
\end{equation}

\noindent where $k_{\rm B}$ is the Boltzmann constant, and $\rho$,
$\phi$, and $r_{0}$ are parameters of the Morse potential used in
Ref. \cite{44} to describe the interatomic interaction potential in
metal, $U(r)=\rho \left[ e^{-2\phi (r-r_{0})}-2e^{-\phi
(r-r_{0})}\right]$.

Note that we consider the thermal expansion of a nanoparticle, by
assuming that it is free. However, the nanoparticle is embedded in
the silica matrix. Respectively, since the volume thermal expansion
coefficient for silica is considerably smaller
(1.65$\times$10$^{-6}$~K$^{-1}$ for fused silica) than one for
copper (5.1$\times$10$^{-5}$~K$^{-1}$), it seems at first glance
that the silica host matrix would block the expansion of a
nanoparticle. However, our procedure of fabrication of Cu/SiO$_{2}$
composite samples is such that the formation of Cu nanoparticles in
silica occurs at the temperature 1473~K, which is considerably
higher than the maximum temperature used in our optical measurements
(460~K). So, the sizes of a nanoparticle and a hosting cavity of the
silica matrix are equal only at the highest temperature, i.e., at
1473~K. After annealing at 1473~K, the samples were cooled down to
room temperature. At cooling, both the nanoparticle and the hosting
cavity contracted. But, due to a considerable difference of the
coefficients of thermal expansion, the copper nanoparticle
contracted considerably stronger, than the hosting cavity did. Thus,
at any temperature lower than 1473~K, including the entire
temperature range 293--460~K used in our experiments, the copper
nanoparticle size is smaller than one of a hosting cavity of the
silica matrix. Therefore, we can conclude that the copper
nanoparticles in our experiments expanded freely, i.e. the matrix
did not affect the thermal expansion of
\mbox{nanoparticles.}\looseness=1

At last, the third mechanism of the temperature dependence of the SP resonance
can be the temperature dependence of the dielectric permittivity of the host
matrix $\varepsilon _{m}(T)$. Indeed, the reference data \cite{45}
show that the permittivity of silica increases with the
temperature. It is seen from Eq. (\ref{eq12}) that such
temperature-induced increase of $\varepsilon _{m}$ would lead to a red shift of the SP
resonance as well. Thus, summarizing the above arguments, we obtain
expressions given below that explain the temperature dependences of the
energy and the width of the SP resonance in metal (copper, in particular)
nanoparticles in the silica host matrix:
\begin{equation}
\omega_{sp}=\sqrt{\frac{\omega_{p0}^{2}}{\left(1+\varepsilon_{ib1}+2\varepsilon_{m}(T)\right)
\left(1+\beta (T)\Delta T \right)}-\gamma^{2}(T)}  ,
\label{eq15}%
\end{equation}\vspace*{-9mm}
\begin{equation}
\gamma (T)=\gamma _{\infty}(T)+A\frac{v_{\rm F}}{R(T)}.
\label{eq16}%
\end{equation}
Here, the dependences $\gamma _{\infty}(T)$, $R(T),$ and $\beta (T)$ are
given by Eqs. (\ref{eq8}), (\ref{eq13}), and (\ref{eq14}),
respectively.

\begin{figure}[b]
\vskip-3mm
\includegraphics[width=4.7cm]{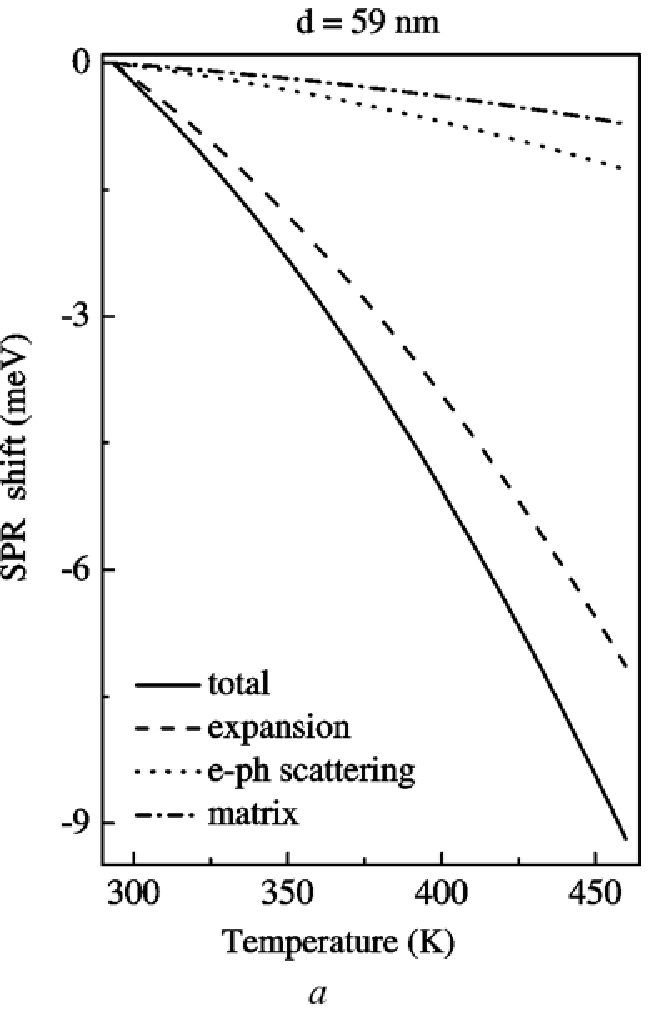}\\[2mm]
\includegraphics[width=4.7cm]{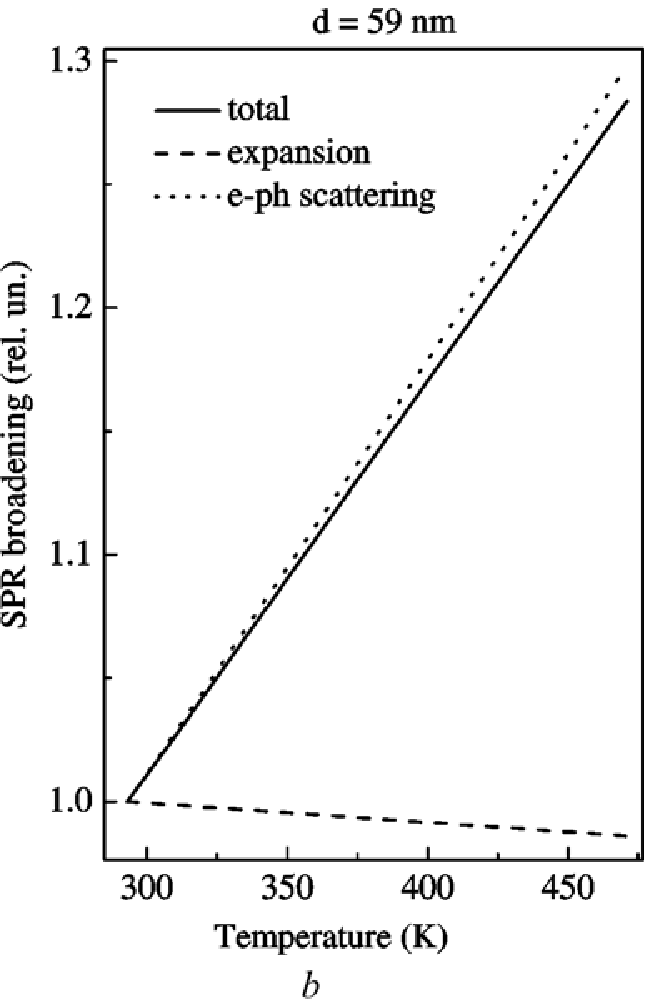}
\vskip-3mm\caption{Calculation of the contributions of different
mechanisms to the temperature-induced ({\it a}) shift and ({\it b})
broadening (the ratio $\gamma (T) / \gamma _{293}$) of the SP
resonance in a 59-nm copper nanoparticle in silica. Solid lines mark
the total shift or broadening, dashed lines -- contribution of the
thermal expansion of a nanoparticle, dotted lines -- the
electron-phonon scattering in the nanoparticle, dash-dotted lines --
the temperature dependence of the dielectric permittivity of the
host matrix}\vspace*{1mm}
\end{figure}

\section{Comparison of Experiment and Theory for~Cu Nanoparticles. Discussion}

In this section, we use the theory outlined in the previous section
to rationalize the experimental temperature dependences of the SP
resonance energy and bandwidth for copper nanoparticles in silica.
We calculate the absorption spectra of Cu nanoparticles in silica
at various temperatures by Eq. (\ref{eq1}). The temperature
dependence of the silica host matrix was taken from Ref. \cite{45}. To
take the temperature dependence of the dielectric
permittivity of a nanoparticle into account, we firstly calculate the contribution of
interband transitions to $\varepsilon$ of a nanoparticle as
\begin{equation}
\varepsilon _{ib}(\omega)=\varepsilon _{\infty}(\omega)-\varepsilon
_{{\rm D},\infty}(\omega),
\label{eq17}%
\end{equation}
where $\varepsilon _{\infty}(\omega)$ is the dielectric permittivity
of copper taken from Ref. \cite{46}, and $\varepsilon _{{\rm
D},\infty}(\omega)$ is the Drude term for bulk copper (contribution
of free electrons) calculated by Eq. (\ref{eq3}). In calculations,
we used $\omega _{p} = 7.6$~eV and $\gamma _{\infty} = 0.09$~eV
\cite{42}. Then we calculated the temperature-dependent dielectric
permittivity of a nanoparticle as follows:
\begin{equation}
\varepsilon(\omega ,T)=\varepsilon _{ib}(\omega)+\varepsilon _{\rm
D}(\omega , T).
\label{eq18}%
\end{equation}
Here, $\varepsilon _{\rm D}(\omega , T) $ was calculated by
Eq.~(\ref{eq3}). In calculations of $\varepsilon _{\rm D}(\omega ,
T),$ (1) the temperature-dependent bulk plasmon frequency was
calculated as $\omega _{p} =$\linebreak $= \omega
_{p0}/\sqrt{1+\beta (T)\Delta T}$, where $\omega _{p0}$ = 7.6~eV,
and the volume thermal expansion coefficient was calculated by Eq.
(\ref{eq14}); (2) the temperature-dependent damping constant $\gamma
(T)$ was calculated by Eq.~(\ref{eq16}), where $\gamma _{\infty}(T)$
was calculated by Eqs.~(\ref{eq8}) and (\ref{eq9}), and $R(T)$ was
determined from Eq.~(\ref{eq13}). In calculations, we also used
$v_{\rm F}=1.57\times 10^{8}$~cm/s for the Fermi velocity in bulk
copper, $A = 0.11$ was estimated from the fitting of the
experimental size dependence of the SP resonance bandwidth at $T_{0}
= 293$~K, and the parameters of the Morse potential $\rho =
0.3287$~eV, $\phi = 13.123$~nm$^{-1}$, and $r_{0} = 0.28985$ nm were
taken from Ref.~\cite{47}.

The evolution of the calculated absorption spectrum of 59-nm copper
nanoparticles in silica with increasing temperature is presented in
Fig.~2~{\it b}. It is seen that an increase of the temperature has
to lead to the red shift and the broadening of the surface plasmon
band. This trend is similar to one observed experimentally. To check
quantitatively our above assumptions of the physical mechanisms of
these two phenomena, we also calculated the temperature dependences
of the shift and the broadening of the SP resonance in copper
nanoparticles. To compare the respective experimental and calculated
dependences properly, we obtained the theoretical values of SP
energy and bandwidth by the decomposition of the calculated spectra
to the basic spectral Lorentzian contours, since the corresponding
experimental values we determined from the decomposition of the
experimental spectra. The obtained calculated temperature
dependences of the SP resonance energy and bandwidth are presented
in Fig.~3. The respective experimental dependences are shown in this
figure as well. It is seen that a quite good agreement of the
experimental and calculated dependences takes place. This fact
proves the correctness of our \mbox{theoretical model.}\looseness=1

As was noted above, the theoretically calculated dependences contain
contributions of three mechanisms. Those are the electron-phonon
scattering in a nanoparticle, thermal expansion of a nanoparticle,
and temperature dependence of the dielectric permittivity of the
host matrix. It is important to know the relative contribution of
each mechanism to the total temperature effect. To do so, we
performed such calculations for the temperature dependences of the
SP resonance energy and width. The results for 59-nm Cu
nanoparticles in silica are presented in Fig.~4. It is seen from
Fig.~4,~{\it a} that the thermal expansion is the dominant mechanism
of the temperature-induced red shift of the SP resonance in copper
nanoparticles. Contributions of the increase of both the
electron-phonon scattering rate and the dielectric permittivity of
the host matrix with increasing temperature are close to each other
and are quite small as compared with the contribution of the thermal
expansion. Meanwhile, the dominant mechanism of the
temperature-induced broadening of the SP resonance in copper
nanoparticles is the electron-phonon scattering, which is seen from
Fig.~4,~{\it b}. Thermal expansion of a nanoparticle leads to a very
small decrease of the SP resonance width. However, this decrease is
negligibly small and can be neglected. The temperature dependence of
the dielectric permittivity of the host matrix does not affect the
SP resonance width. Note that the good agreement of experiments with
the above-outlined theory takes place as well for the temperature
dependence of the SP resonance in silver nanoparticles \cite{36}.
This proves the similarity of the physical mechanisms of the
observed temperature dependences for SP resonances in copper and
silver nanoparticles, i.e. the thermal expansion of a nanoparticle
as a cause of the red shift and the electron-phonon scattering as a
cause of the broadening of the SP resonance. It is seen from
Fig.~3,~{\it a} that the red shift rate of the SP resonance with
increasing temperature is the same for Cu nanoparticles of all
studied sizes. This indicates that the volume thermal expansion
coefficient is size-independent in the studied range of Cu
nano\-par\-ticle~sizes.\looseness=1

\section{Conclusions}
The temperature dependences of the SP resonance energy and width in
copper nanoparticles with mean sizes of 17, 35, 48, and 59 nm
embedded in the silica glass host matrix are studied in the
temperature range 293--460 K. We observed that, as the temperature
increases, the red shift and the broadening of the SP resonance
occur. A theoretical model including the phenomena of
electron-phonon scattering in a nanoparticle, the thermal expansion
of the nanoparticles, and the temperature dependence of the
dielectric permittivity of the host matrix is considered to
rationalize the observed temperature behavior of the SP resonance in
copper nanoparticles. As the temperature of a particle increases,
the nanoparticle volume increases, and the density of free electrons
decreases. The lower electron density leads to the lower plasma
frequency of electrons and subsequently to the red shift of the SP
resonance. The rate of electron-phonon scattering increases with the
temperature. This leads to an increase of the damping constant of
plasma oscillations and, as a result, to the red shift of the SP
resonance. The dielectric permittivity of silica increases with the
temperature, which leads to a red shift of the SP resonance as well.
The results of calculations of the temperature dependences of the SP
resonance energy and width are in very good agreement with the
respective experimental data, which proves the validity of the used
theoretical model. It is shown that the thermal expansion of copper
nanoparticles is the dominant mechanism of the temperature-induced
red shift of the SP resonance in copper nanoparticles. Meanwhile,
the dominant mechanism of the SP resonance broadening with
increasing temperature is the electron-phonon scattering in Cu
\mbox{nanoparticles.}\looseness=1

\vskip3mm {\it The author thanks Dr. A.A. Alexeenko for samples of
Cu/SiO$_{2}$ nanocomposite, Dr. A.M. Dmytruk for TEM measurements,
and Prof. I.M. Dmitruk for fruitful discussions.}

\end{document}